\documentclass[preprintnumbers,showpacs, showkeys,
 twocolumn,secnumarabic,amssymb,amsmath,amsfonts,aps, pra]{revtex4}

 \usepackage{graphicx}

\begin{document}

 \newcommand{\non}{\nonumber}
 \newcommand{\be}{\begin{equation}}
 \newcommand{\ee}{\end{equation}}
 \newcommand{\bq}{\begin{eqnarray}}
 \newcommand{\eq}{\end{eqnarray}}
 \newcommand{\lps}{\langle}
 \newcommand{\rps}{\rangle}

 \title{Nonclassical dynamics of Bose condensates in an optical lattice
	in the superfluid regime}

 \date{\today}

 \author{Roberto Franzosi}
 \email{franzosi@fi.infn.it}
 \affiliation{Dipartimento~di~fisica~Universit\`a~di~Firenze~and~CNR-INFM,
~Via~Sansone~1,~I-50019~Sesto~Fiorentino,~Italy.}

 \begin{abstract}
 A condensate in an optical lattice, prepared in the ground state of the
 superfluid regime, is stimulated first by suddenly increasing the optical
 lattice amplitude and then, after a waiting time, by abruptly decreasing
 this amplitude to its initial value.
 Thus the system is first taken to the Mott regime and then back to the
 initial superfluid regime.
 We show that, as a consequence of this nonadiabatic process, the system falls
 into a configuration far from equilibrium whose
 superfluid order parameter is described in terms of a particular superposition
 of Glauber coherent states that we derive.
 We also show that the classical equations of motion describing the time
 evolution of this system are inequivalent to the standard discrete nonlinear
 Schr\"odinger equations. By numerically integrating such equations with
 several initial conditions, we show
 that the system loses coherence, becoming insulating.

 \end{abstract}
 \maketitle


\section{Introduction}
Nowadays Bose-Einstein condensates (BECs) represent one of more
powerful and versatile testing grounds for low-energy modern physics
in which experimental tests on quantum computation \cite{Mandel_Nature425},
many-body physics \cite{Mandel_PRL91}, superfluidity \cite{Burger_PRL86},
the Josephson junction effect \cite{Cataliotti_Science293}, atom optics
\cite{Ottl_PRL95}, and quantum phase transitions \cite{Greiner_Nature415}
can be performed.
Condensates can be put into interaction with each other or manipulated
by means of optical lattices (OLs), which are periodic trapping
potentials generated by standing laser waves.
By raising the amplitude of the OL, a condensate
loaded therein is fragmented into an array of interacting condensates.
By adjusting the laser amplitudes, the system is taken into
different regimes. The superfluid regime is obtained with weak optical
potentials (OPs), where the kinetic energy dominates over the interacting
one, and the atoms hop from one well to another.
The opposite -quantum- regime, is generated by strong OPs
that suppress the tunneling of the atoms between the wells.

In this paper, we consider a one-dimensional gas. This is prepared by
use of a transverse harmonic confining potential that tightly confines
the atoms so that their motion, in the transverse direction, is
limited to the zero point.
Along the longitudinal direction, a further harmonic potential weakly
confines the atoms, and an OP is switched on.
For large enough laser amplitudes, the condensate splits into components
tightly confined at the minima of the effective potential.
In what follows, we imagine abruptly adjusting the amplitude of the
longitudinal laser at two instants, in order to first take the system
from the superfluid regime to the quantum one and then to take
it back to the superfluid regime.
In this way the system is taken to a nonequilibrium state (NES) that is
described in terms of a particular superposition of Glauber coherent states (CSs).
The latter combination is derived in the following.
Although the system is taken back to a weak OP regime, as in the
superfluid case, this NES follows a nonclassical
dynamics.
In fact, we show that the equations of motion
for the system's order parameter
are inequivalent to the
discrete nonlinear Schr\"odinger equation.
The present study is in some sense complementary to former work~\cite{Altman_PRL89,Altman_PRL95}, it continues the work
in~\cite{Buonsante_JPB37_1}, and it is motived by recent experiments
as in Refs.~\cite{Greiner_Nature415,Orzel_Science291,Kinoshita_Science305},
where BECs are manipulated in an OL. Furthemore, the nonadiabatic procedure
we describe, can be straightforwardly realized in a real experiment, and
then the dynamics of the NES, a superposition of the canonical CSs can be
directly observed,
e.g., by displacing the condensates with respect to the harmonic trap center,
as in the experiment of Ref.~\cite{Cataliotti_Science293}, and
by observing the oscillations of the center of the atomic density
distribution.

\section{The model}
The quantum dynamics of an array of condensates in a deep enough OL,
can be described by the
Bose-Hubbard (BH) model \cite{Jaksch_PRL81}.
Let  $V_{H} ({\bf r})=\Sigma^3_{j=1} m \Omega^2_j
r^2_j /2$ be the harmonic trapping potential and
let $V_{L} ({\bf r})=  \hbar^2 \omega^2  \sin^2(k r_1)/(4 E_r)$ 
be the one-dimensional OP, where $k$
is the laser wavelength, $E_r =\hbar^2 k^2/(2 m)$ 
is the recoil energy, and $\omega$ is the angular frequency
of the parabolic approximation of $V_L$ at each minimum.
Then, the BH Hamiltonian, written in terms of the
boson operators $a_{ j}$ and $a^{+}_{ j}$ that, respectively, annihilate 
and create atoms at the $j$ site of the lattice, reads
 \be 
 H=  \sum_{ i}[\, U n_{ i}(n_{ i}-1) + \lambda_{ i} n_{ i} ] -
 {\frac{T}{2}} \sum_{\lps{ ij} \rps}  \,
 \left (a^{+}_{ i} a_{ j} + h.c. \right ) \, ,
 \label{BHH}
 \ee
 where the operators $n_{ i}=a^{+}_{ i} a_{ i}$ count the number of bosons 
 at the $i$ site, and the boson operators satisfy the standard commutation 
 relations $[ a_{i}$, $a^{+}_{j}] = \delta_{i,j}$. The indices $i,j \in 
 \mathbb{Z}$ label the local minima ${r_1}_{i},{r_1}_j$, where  ${r_1}_\ell = \pi 
 \ell/k$, of $V({\bf r})= V_L ({\bf r}) + \Sigma^3_{\ell=2} m \Omega^2_\ell 
r^2_\ell /2$ along the lattice.
As the BH model describes a closed system, the total number of bosons
$N=\Sigma_j n_j$ is a conserved quantity.
Within the Gaussian approximation, the Hamiltonian parameters have the following
expressions in terms of the trapping potentials and of the optical one
(see \cite{Buonsante_JPB37_1}).
$U:= a_s \Omega_0 \sqrt{m \hbar \Omega_0 /(2 \pi)}$ is the
strength of the on-site repulsion, in which we have set $\Omega_0
 = \sqrt[3]{\omega \Omega_2 \Omega_3}$.
The latter approximation is suitable in the
limit of tight confinement of BECs in every well. In fact, in this case the
spatial width of each trapped condensate does not depend, in the first
approximation, on the number of atoms in the well, and the condensate
wave function in every potential minimum is well approximated by a
Gaussian~\cite{MSTnjp}.
The site external potential is $\lambda_{ j}:=\epsilon (\omega)
 +{ j}^2 \pi^2 \hbar^2 \Omega^2_1/(4 E_r)$, where $\epsilon (\omega)
= \hbar (\omega + \Omega_2 + \Omega_3)/2$, and
 \be
 T(\omega)= \frac{\hbar^2 \omega^2}{4 E_r}
 \left[ \frac{\pi^2}{2} - 1 + \frac{2 E_r}{\hbar\omega} -
 e^{- \frac{2 E_r}{\hbar \omega}} \right] 
 e^{- \frac{\pi^2\hbar \omega }{8 E_r} } \,
 \label{hopping}
 \ee 
is the tunneling amplitude between neighboring sites.
It is worth stressing that the Gaussian approximation is not essential
for the use of the BH model, but it is useful in order to  derive an
analytic estimation of the Hamiltonian parameters.

When the OL amplitude is weak enough so that
$T/U \gg 1$, the ground-state configuration of Hamiltonian (\ref{BHH})
admits a factorization into a product of site states that catches the
superfluid nature of the system.
The system's order-parameter dynamics near the ground state
can be studied by a time-dependent variational principle (TDVP)
\cite{Zhang_RMP62}.
Following the TDVP method, we describe the system in terms of the
trial state $| \Psi \rangle ={\rm exp}(iS/\hbar) \Pi_i |z_i \rangle$,
which contains a product of site Glauber CSs. In fact, in Ref.~\cite{Zwerger_JOB5}
it is shown that, in the limit $U=0$, $N,M \to \infty$
at fixed density $N/M$, the ground state of Hamiltonian (\ref{BHH}) with
$M$ wells, is indistinguishable by a product of local coherent states. Thus,
in the strong hopping regime, such a state should still be a good approximation.
Here
 \be
 |z_i \rangle : = {\rm e}^{- \frac{1}{2}|z_i |^2}
 \sum^{\infty}_{n=0} \frac{z_i^n}{n!} \,
 (a^{\dagger}_i )^n |0 \rangle \, 
 \label{CS}
 \ee
have the defining equation $a_i | z_i \rangle = z_i| z_i  \rangle$,
and the $z_i$ are complex numbers. 
The equations of motion for the $z_j$ dynamical variables
are derived by a variation with
respect to $z_j$ and $z^*_\ell$ of the effective action $S=\int dt 
(i \Sigma_j \dot{z}_j z^*_j -{\cal H})$ that is associated with the classical
Hamiltonian ${\cal H}( Z, Z^*):= \langle Z| H |Z \rangle$,
where $|Z \rangle = \prod_j |z_j \rangle$. 
Hence, the
$z_j= \langle \Psi| a_j |\Psi \rangle$
represent the classical canonical variables of the effective Hamiltonian
${\cal H}$ and satisfy the Poisson brackets $\{z^*_j,z_\ell  \} = i 
\delta_{j \ell} /\hbar$.
The classical Hamiltonian is
\be
{\cal H} = \sum_{j} \Bigl [ U |z_j|^4  +
\lambda_j |z_j|^2 - \frac{_{T}}{^2}
\left (z^{*}_j z_{j+1} +
\ {\rm c.c.} \ 
\right ) \Bigr ] \,  ,
\label{HS}
\ee
where $j$ and $j+1$ run on the chain sites, and the
following equations of motion result:
\bq
 i\hbar {\dot z}_j &=& (2U|z_j|^2 +\lambda_j ) z_j
-\displaystyle \frac{_{T}}{^2} (z_{j-1}+z_{j+1}) \; ,
\label{CEM-1}
\eq
together with the complex conjugate equations. The constraint
on the total number of bosons is now satisfied on average,
the quantity $N = \sum_j |z_j|^2$ being conserved.

Equations (\ref{CEM-1}) are the discrete version of the Gross-Pitaevskii 
equation~\cite{MSTnjp},
and the corresponding superfluid ground state is approximately
given by the discrete Thomas-Fermi solution
 \be
 z_j = \sqrt{\frac{N}{M^\prime}-\frac{(\lambda_j - \bar{\lambda})
 }{2 U}}\  e^{i \phi} \, ,
\label{T-F}
 \ee
where
$M^\prime=min(M,q)$ and $q$ in turn is the maximum
integer such that $2UN + q (\bar{\lambda} - \lambda_q ) \geq 0$, and
$\bar{\lambda}= \Sigma_{_{|j|\in  {I_{_{M^\prime}}}}} \lambda_j/ M^\prime$.

\section{Dynamics}
After having prepared the system in the superfluid ground-state configuration
(\ref{T-F}), it is taken to the Mott regime by abruptly increasing the OP
depth and, after an adjustable time
$\tau$, it is carried back to the superfluid regime by suddenly 
decreasing the OP depth to the original value.
Our goal is to derive the equations of motion that describe the dynamics
of the system after the latter decreasing of the OP
amplitude. We will show that these equations of motion are more complicated
than the standard ones recalled in Eq. (\ref{CEM-1}), and inequivalent
to these.
Furthermore, as we said above, the superfluid ground state is well
approximated by a product $\prod_i |z_i\rps$ of CSs.
Hence, the condensate in each site $i$ is described by a CS
$|z_i\rps$, that is, a semiclassical state.
On the contrary, we shall show that, after this
stimulation, the system will be described by a
product of integrals of site CSs. This means that the
semiclassical nature of the site states is partially destroyed during the
intermediate quantum regime, in spite of the system being in a superfluid
regime.

Following the procedure above, at the time $t=0$ the amplitude
of the OP is suddenly increased by varying $\omega$
from its initial value $\omega_0$ to $\omega = \omega_1 \gg
\omega_0$. Since the tunneling amplitude $ T(\omega)$ in Eq. (\ref{hopping})
is dominated by the exponential term  $\exp \{- {\pi^2\hbar \omega }/
{8 E_r}\}$, it will result in $T(\omega)/T(\omega_0)\to 0$ for
$\omega \to \omega_1$.
Meanwhile, also $U$ and $\lambda_j$ are modified when changing the
OP amplitude, but their dependence on $\omega$ is much less
dramatic. In fact, we have
$U(\omega_1) = U \sqrt{\omega_1/ \omega_0}$
and $\lambda_j (\omega_1) = \lambda_j + \hbar (\omega_1 - \omega_0)/2 $.
In order to apply the sudden approximation when
the potential amplitude is varied, that is, for $0<t<\tau_b $
with $\tau_b \ll \tau$, the jump in the potential depth must be fast compared
with the tunneling time between neighboring wells, but slow enough so
that no excitations are induced in each well, that is $ 2 \pi/\omega,
2 \pi/\Omega_2, 2 \pi/\Omega_3 \ll \tau_b \ll \hbar/T(\omega_0) $.

For $t>\tau_b$ (we will assume $\tau_b = 0$ from now on), the system enters
into the Mott regime and the classical description of the system
dynamics is no longer allowed; thus we resort to the quantum one.
The appropriate dynamics is described
by the Schr\"odinger equation with Hamiltonian (\ref{BHH})
in which we have to set $T=0$, $U=U(\omega_1) =:\tilde U$ and
$\lambda_j = \lambda_j (\omega_1) =: \tilde  \lambda_j$.
The quantum time evolution of the initial state (\ref{CS})
is
\begin{equation}
 |(t)\rangle
 = \prod_{i\in I_{_M}} e^{ - \frac{|z_i|^2}{2}} \sum_{n_i=0}^{+\infty}
 \frac{[z_i \ \nu_i(t)]^{n_i}}{\sqrt{n_i !}} e^{-i n^2_i u(t)}
 | n_i \rangle \, ,
 \label{CST}
\end{equation}
where $\nu_i(t):= \exp [i/\hbar (\tilde U+\tilde \lambda_i) t]$,
$u(t) =  \tilde U t/ \hbar$, and the $z_j$ are those defined in~(\ref{T-F}). 
We want to stress that, although the factorization
of the state vector still holds,
the term $\exp [ -i n^2_i u(t)]$ in Eq. (\ref{CST}) breaks the
CS structure of the initial state (\ref{CS}), and
$a_{j}|(t)\rangle \neq \alpha(t) |(t)\rangle$.
By direct calculation, one can easily verify the following relations:
\be
 \begin{split}
 &\langle(t) |a^+_{j} a_{j}|(t)\rangle = |z_j |^2 \, , \\
 &z_j(t):=\langle(t) | a_{j}|(t)\rangle = z_j
 e^{  \frac{i}{\hbar} \tilde \lambda_j t  - 
 i|z_j|^2 \sin [2u(t)]  }
 \\
 &~~~~~~~~~~~~~~~~~~~~~~~~~~~~~~~ 
 e^{-2 |z_j|^2 \sin^2 [u(t)] } \, , \\
 &\langle(t) |a^+_{j+1} a_{j}|(t)\rangle  = 
 z^*_{j+1} z_j 
 e^{-2 (|z_j|^2 + |z_{j+1}|^2)\sin^2 [u(t)] }
 \\
 & ~~~~~~~~~~~~~~~~~~~
e^{ \frac{i}{\hbar} 
  (\tilde \lambda_j - \tilde \lambda_{j+1})t -
  i (|z_j|^2 - |z_{j+1}|^2)\sin [2u(t)] 
  }  \, .
 \label{ics}
 \end{split}
\ee
The system shows the characteristic scenario of phase collapse and
revivals, observed in many BEC
systems~\cite{Greiner_Nature419,Sinatra_Castin_EPJD8}.
For $0< t <\tau$, the wells' populations $\langle(t) |a^+_{j}
a_{j}|(t)\rangle$ do not change, whereas site wave functions $z_j(t)$
are dynamically active.
The modulus of $z_j(t)$ is a periodic function of $t$, whose revival time is
$T_m = \pi \hbar/ \tilde U$.
The phase of the wave function $\varphi_j := \arg 
 [z_j(t)/\vert z_j(t) \vert] = \tilde \lambda_j t/{\hbar}   - 
 |z_j|^2 \sin [2 u(t)]$ of the site $j$, is driven by the three time scales
$2 T_m$, $T_{\tilde \lambda_j} = 2 \pi \hbar /\tilde \lambda_j$, and,
for the sites where $|z_j|^2 >2 \pi$, $T_{z_j}$, the solution of the equation
$|z_j|^2 \sin^2 [u(t+T_{z_j})] = |z_j|^2 \sin^2 [u(t)] + 2 \pi$.
Moreover, the site-dependent external potentials $\tilde \lambda_j$
induce a dephasing between the wave functions of near sites:
$\varphi_j - \varphi_{j+1} = -
(\pi \hbar \Omega_1/2)^2 (2j+1)/E_r - (\vert z_j \vert^2 -
\vert z_{j+1} \vert^2) \sin^2 [u(t)]$.
Such dephasing leads the system out of the ground-state configuration.

After a time $\tau$, the system is taken back to the superfluid regime,
$T/U \gg 1$, by abruptly decreasing the OP depth (in 
a time of order $\tau_b \approx 0$) to its initial value.
The $t=\tau$ initial state
 \be
 |(\tau)\rangle = 
 \prod_{j\in I_{_M}} {\cal E}_j \sum_{n_j=0}^{+\infty}
 \frac{[z^1_j]^{n_j}}{\sqrt{n_j !}} e^{-i n^2_j u_1}
 | n_j \rangle~~~~ \, 
\label{is3}
 \ee
given by Eq. (\ref{CST}), where ${\cal E}_j = \exp \{ - |z_j|^2/2 \}$, 
$z^1_j = z_j \nu_j(\tau)$, and $u_1 = u(\tau)$, by  the identity 
 \[
 \lim_{\epsilon \to 0^+}
 \int^\infty_{-\infty} d x \exp 
 [-(p + \epsilon) x^2 - in x] = \exp [- n^2/(4 p)] \sqrt{ \pi/p}  \, ,
 \]
with $p=-i/(4u_1)$, can be rewritten as the superposition of product
states of CSs at each site,
  \be
  |(\tau)\rangle = \!
  \prod_{j\in I_{_M}}  \!
  \int^\infty_{-\infty}  \frac{d x_j}{2 \sqrt{\pi u_1}}
  e^{-i \pi /4 } 
  e^{i x^2_j / (4 u_1)} |  z^1_j e^{- i x_j }\rps \, ,
  \label{sup}
  \ee
where the states labeled by $z^{\prime\prime}_j=z^1_j e^{- i x_j }$ 
are the normalized CSs of Eq. (\ref{CS}) with $z_j = 
z^{\prime\prime}_j$.

The time evolution for $t>\tau $ of the {\em mean-field} state (\ref{sup}) can be derived
within the TDVP picture, in a way similar to that previously described,
by resorting to a {\em suitable superpositions of Glauber CSs}.
Thus we introduce the trial
state $ | \Psi \rangle ={\rm exp}(iS/\hbar) |Z\rangle_v$, where
$|Z \rangle_v := \Pi_i |z_i , u_1\rangle_v$ is written in terms of the states
$|z_i , u_1\rangle_v$ that in turn are superposition of the standard Glauber
ones as
\be
 |z_i , u_1 \rangle_v : = \int^{\infty}_{-\infty} d x f(x,u_1) |  z_i
 e^{-i x} \rangle \, .
 \label{vCS}
\ee
Here $|  z_i e^{-i x}\rangle$ are the standard CSs given
in (\ref{CS}), and 
\be
 f(x,u_1)= \frac{1}{2 \sqrt{\pi u_1}}
 e^{-i \pi /4 } 
 e^{i x^2 / (4 u_1)} \, .
 \label{f}
\ee
A remark about the trial state that we have chosen is in order.
The TDVP method provides the best approximation to the true state within
the restricted set of states caught by the trial one; thus, in general,
we do not know what superposition of the canonical CSs gives a class of
states broad enough to obtain a good approximation of the true state.
However, in this case, the form of the superposition (\ref{vCS}) is suggested
by the fact that it includes the initial condition (\ref{sup}).
Furthermore, in the limit $\tau \to 0$,
that is, $u_1 \to 0$, $|z_i , u_1\rangle_v \to |z_i \rangle$ and, in this way,
the canonical CSs (\ref{CS}) and the standard dynamics (\ref{CEM-1}) are recovered.
From now on we drop the explicit dependence on $u_1$ in $|z_i,u_1 \rangle_v$.
The scalar product between the $|z_\ell \rangle_v$ states is defined as
$_v\langle z_j  |z_\ell \rangle_v :=
\int d x d y f^*(x,u_1) f(y,u_1) \langle z_j e^{-i x} 
|z_\ell e^{-i y}\rangle$ (here the integrations over $x$ and $y$
run from $-\infty$ to $\infty$); thus, from the definition (\ref{vCS}) and
by the normalization of the Glauber CSs $\langle \alpha | \alpha^\prime \rangle
= \exp [\alpha^* \alpha^\prime -1/2(|\alpha|^2 + |\alpha^\prime|^2)]$, the
following identities can be checked by direct calculation:
\be
 \begin{split}
 _v\langle z_j  |z_j \rangle_v =& 1 \, , \\
 _v\langle z_j| i \hbar \partial_t  |z_j \rangle_v =&
 \frac{i \hbar}{2} (z^*_j \dot{z}_j - \dot{z^*}_j z_j)  \, , \\
 _v\langle z_j| a^\dagger_j a^\dagger_j a_{j} a_j|z_{j} \rangle_v =&
 |z_j|^4 \, , \\
 _v\langle z_j| a^\dagger_ja_j|z_{j} \rangle_v =&
 |z_j|^2 \, , \\
 _v\langle z_j| a^\dagger_j a_{j+1} |z_{j+1} \rangle_v =&
z^*_j z_{j+1} 
 e^{ -2 (|z_j|^2 + |z_{j+1}|^2)\sin^2 [u_1] } \\
&~~~~~~~~~
 e^{  i (|z_j|^2 - |z_{j+1}|^2)\sin [2u_1]
} \, .
 \label{vCSp}
 \end{split}
\ee
Following the TDVP procedure, we require the trial state to be a
solution of the weaker form of the Schr\"odinger equation,
$$
\langle \Psi | i \hbar \partial_t - H |\Psi \rangle = 0 \, ,
$$
where $\partial_t$ is the time derivative and $H$ is the BH Hamiltonian
(\ref{BHH}). From the latter equation, we obtain 
$$
\dot{S} = i \hbar \, _v\langle Z|
\partial_t |Z\rangle_v -\, _v\langle Z| H |Z\rangle_v \, ,
$$
and, from this and the relations in (\ref{vCSp}), we get
\be
{S} =\int d t [i \hbar  \sum_j \frac{1}{2} (z^*_j \dot{z}_j - \dot{z^*}_j z_j) -
 {\cal H} (Z,Z^*)] \, .
\label{Action}
\ee
The effective classical Hamiltonian ${\cal H} (Z,Z^*) :=
\, _v\langle Z| H |Z\rangle_v$, can be derived by exploiting the identities
in (\ref{vCSp}) and the result is
\be
 \begin{split}
& {\cal H} (Z,Z^*)  = \sum_j
[U |z_j|^4 + \lambda_j |z_j|^2] + \\
 -& \frac{T}{2} \sum_j[ z^*_j z_{j+1} 
e^{  i (|z_j|^2 - |z_{j+1}|^2)\sin [2u_1]}
 + 
 c.c.]\\
&~~~~~~~~~~~~~~~~~~
 e^{ -2 (|z_j|^2 + |z_{j+1}|^2)\sin^2 [u_1] }  \, .
 \end{split}
 \label{evH}
\ee
The variation of the action (\ref{Action}) with respect to $z_j$ and $z^*_j$
brings us to the classical equations of motion
\bq
 \begin{split}
 i\hbar {\dot z}_j
   &= (2U|z_j|^2 +\lambda_j 
) z_j  -\displaystyle \frac{_{T}}{^2}
  \times \\
 &
 \left\{
 \left[ z_{j+1} + z^*_{j} z^2_{j+1} - \frac{1}{2} |z_{j}|^2 z_{j+1} 
 \right] e^{\Phi_{j, j+1}} + \right. \\
 &\left[ z_{j-1} + z^*_{j} z^2_{j-1} - \frac{1}{2} |z_{j}|^2 z_{j-1} 
 \right] e^{\Phi_{j, j-1}} + \\
 & 
 \left.
 - \frac{1}{2} z^*_{j+1} z^2_{j} e^{\Phi_{j+1, j}} 
   - \frac{1}{2} z^*_{j-1} z^2_{j} e^{\Phi_{j-1, j}} \right\}\; 
 \end{split}
 \label{CEM}
 \eq
for $j\in I_{_M}$, where $\Phi_{j, k} = z^*_j z_k - (|z_j|^2 + |z_k|^2)/2$,
and with the complex conjugate equations.
\begin{figure}[t]
\includegraphics[scale=.26]{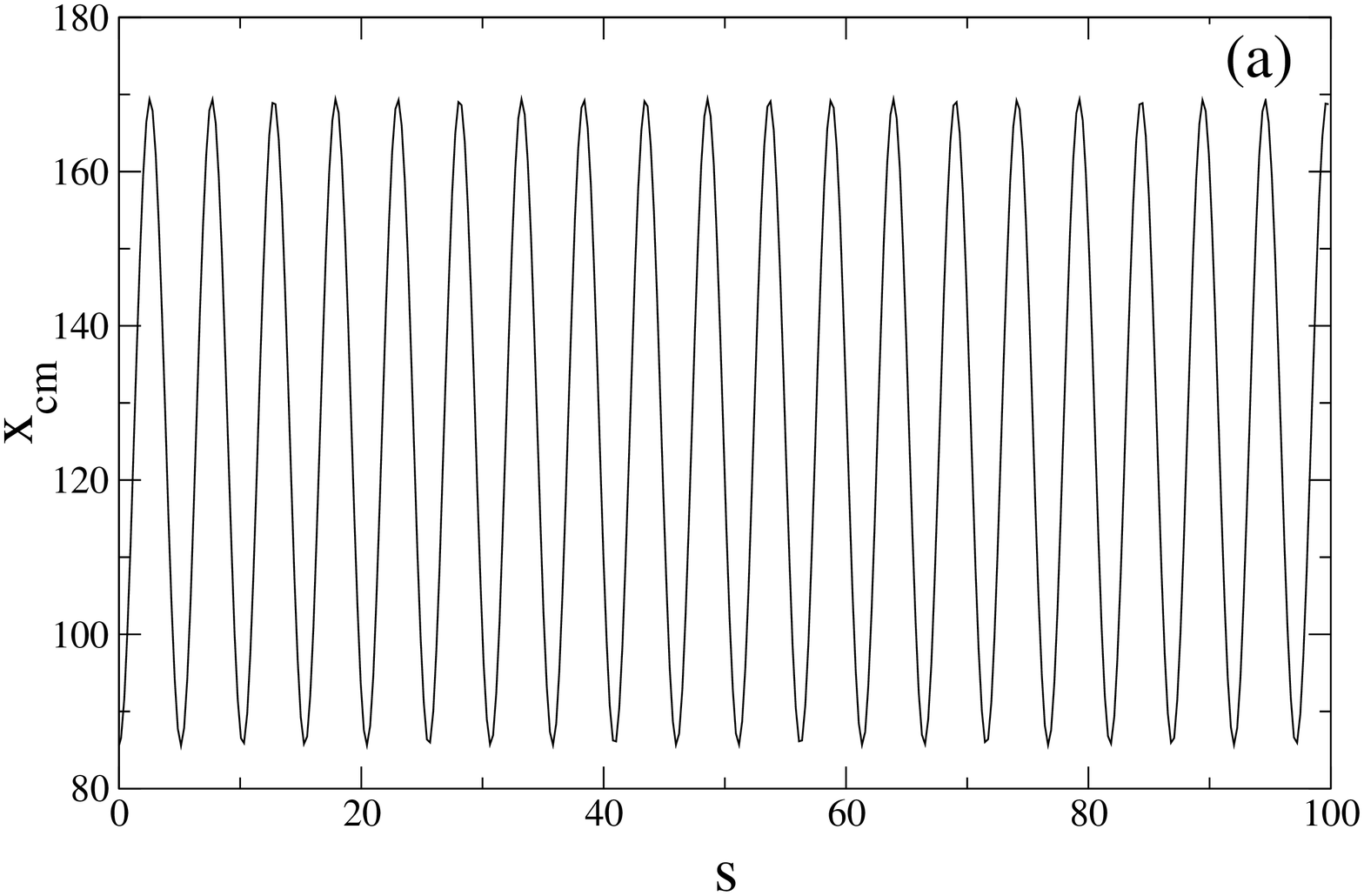}
\includegraphics[scale=.26]{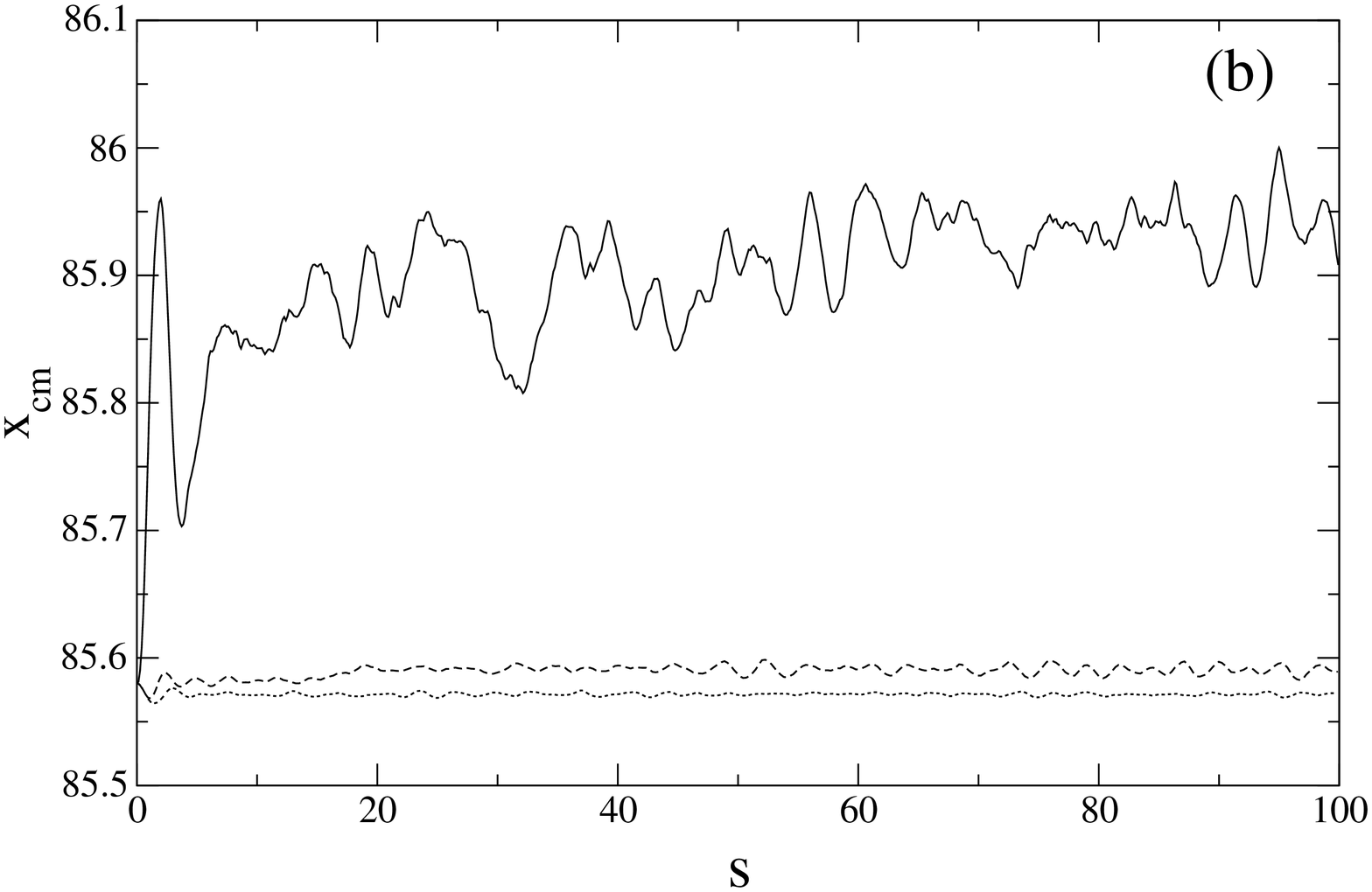}
\caption{Center of the atomic density distribution
$x_{cm} = \sum_j |z_j|^2 j$, as a function
of the rescaled time $s = t \hbar/U$ (both dimensionless).
(a) shows the regular oscillations of the center of the atomic density distribution
of a BEC in a harmonic trap obtained by numeric integration of Eqs. (\ref{CEM-1}).
(b) shows the same quantity obtained by numerical integration
of Eqs. (\ref{CEM}) with $\tau = \pi/10$ (continuous),
$\pi/4$ (dashed), and $\pi/2$ (dotted) lines.}
\label{c-o-m}
\end{figure}
It is worth noting that, in the present case, the dynamical variables
$z_j,z^*_\ell$ are related to the expectation values of the boson
operators in a more complicated form than in the Glauber CS case.
In fact we have $\, _v\lps z_j| a_j |z_j \rps_v$, =
$z_j e^{-i u_1} \exp[|z_j|^2(e^{-i2 u_1}-1)] $.
Despite that, the $|z_j|^2$ still count the number of atoms in each site $j$,
see the fourth of Eqs. (\ref{vCSp}).

\section{Numerical simulations}
We have numerically integrated Eqs. (\ref{CEM-1}) and (\ref{CEM})
on a lattice of $256$ sites, with the initial conditions (\ref{vCSp})
with the $z_j$ given by (\ref{T-F}), and
for the values of $u_1$ corresponding to the waiting times $\tau=\pi/10,\pi/4,\pi/2$.
For a BEC in an OL with $T/(2U N) \approx 0.01$,
in a harmonic trap with $\lambda_j/(2 U) = \alpha (j-128)^2$ and
$\alpha=0.01$,
the standard equations of motion (\ref{CEM-1}) imply a superfluid
dynamics \footnote{These parameters are close to the experimental ones
used in \cite{Cataliotti_Science293} where it has been shown that the
tight-binding approximation describes very well the superfluid dynamics
of BECs in OLs. Those parameters are:
$\hbar^2 \omega^2/(4 E_r) = 3 E_r$, $T/2 \approx 0.07 E_r $, $2 U N\approx
12 E_r$, $N \approx 2 \times 10^5$ atoms, that correspond to the parameter
of the classical dynamics $T/(2U N)
\approx 0.01$.}.
This is shown in Fig.~\ref{c-o-m}~(a), where we plot the regular oscillations
of the center of the atomic density distribution along the chain.
These oscillations have been triggered by displacing the condensates respect
to the harmonic trap center, as in the experiment of Ref.~\cite{Cataliotti_Science293}.
On the contrary, once $\tau \neq 0$,
the nonstandard equations of motion (\ref{CEM}), with the
same initial conditions, entail insulator (dissipative) dynamics for the system.
This is clearly shown in Fig.~\ref{c-o-m}~(b), where we plot the motion
of the system's center of atomic density distribution for $\tau = \pi/10$
(continuous line), $\pi/4$ (dashed line), and $\pi/2$
(dotted line).
\begin{figure}[t]
\includegraphics[scale=.4]{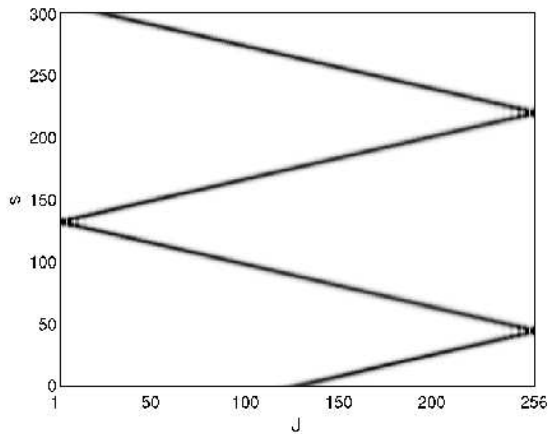}
\caption{Breather excitation obtained by numeric
integration of Eqs. (\ref{CEM-1}) with a Gaussian initial profile.
The breather travels along the chain and bounces at the lattice ends.
After any bounce, the breather is reconstructed.}
\label{SCSsol}
\end{figure}
We also have performed numerical simulations of Eqs.~(\ref{CEM-1})~and~(\ref{CEM})
by choosing initial conditions in the form of wave packets of Gaussian
profile $z^0_j = \sqrt{k} \exp \{ -(j-x)^2/\sigma^2 + i p(j-x) \}$,
where $x=128$ is the initial center of the Gaussian, $p=3 \pi/4$ is the
initial center of mass momenta, $\sigma = 10$ is the width of the Gaussian
profile, and $k=[\sum_j exp(-(j-x)^2/(2\sigma^2))]^{-1}$.
\begin{figure}[t]
\includegraphics[scale=.4]{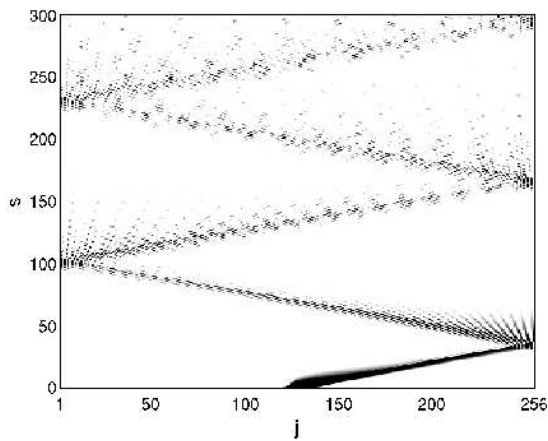}
\caption{Excitation obtained by numeric
integration of Eqs. (\ref{CEM}) with the same initial condition as in
Fig.~\ref{SCSsol}. The value of the waiting time is $\tau = \pi/2$.
This excitation is completely destroyed after a few bounces at the lattice
ends. Thus, in the far-from-equilibrium situation, this excitation loses
stability and it seems to behave like the states of the diffusion regime
that has been identified in~\cite{Trombettoni_PRL86}.}
\label{VCSsol}
\end{figure}
The dynamics of these profiles have been numerically and analytically studied in
Ref.~\cite{Trombettoni_PRL86}, where a dynamical stability
phase diagram for these states was derived. Therein, and also here, the theoretical
configuration where the harmonic trapping is off was considered, and the dynamics takes place on
a finite lattice endowed with reflecting boundary conditions.
Thus, we have performed simulations in the same conditions and we have chosen the
combination of the dynamical parameters $T/(2 U N)=4.17$, which corresponds to the
region of the phase diagram of Ref.~\cite{Trombettoni_PRL86} where (stable) breather
excitations were identified.
In Fig.~\ref{SCSsol} we report the two-dimensional
contour plot obtained by numeric integration of (\ref{CEM-1}) with this
Gaussian initial condition. In Fig.~\ref{SCSsol} is clear that the breather structure
is maintained when the traveling excitation is reflected at the lattice boundaries.
Figure~\ref{VCSsol} shows the same quantity obtained by integration of
Eqs.~(\ref{CEM}) with the same initial condition as in Fig.~\ref{SCSsol}, and with
$\tau=0.5$. Figure~\ref{VCSsol} clearly shows that the initial
excitation, integrated with nonstandard dynamics, pretty soon
loses stability, emitting atoms incoherently at any bounce with the lattice boundary.

\section{Final remarks}
In the present paper we have studied how the dynamics of a superfluid is affected by briefly
bringing the system into the insulating regime.
We have shown that
the system is taken to an excited state, described by
a superposition of product states of Glauber coherent states at each site, which we have
derived.
The classical equations of motion ruling its dynamics have been derived.
Furthermore, we have shown that these classical equations of motion are inequivalent to
the standard discrete nonlinear Schr\"odinger equations that describe the dynamics of
an array of BECs~\cite{Cataliotti_Science293} in the superfluid regime.
By numerically integrating such nonstandard equations with several initial conditions,
we have shown that the system loses coherence, becoming insulating.

The simulations we have performed show that the interplay between classical and quantum dynamics
leads to loss of the coherence properties of the system. In fact, the brief period in the
insulating regimes changes the superfluid wave function, which becomes a superpostion
of product states of the site's coherent states, that is, a product of mean-field states.
Each mean-field state has a complicated distribution of phases at each site that results
from the intermediate quantum dynamics. This distribution of phases leads to a unique
tunneling dynamics described by a complicated hopping term.
By a glance at Eqs. (\ref{CEM}) one can guess that, as a consequence of this unusual
term, the ``effective'' tunneling rate between close sites in the case of
Eqs. (\ref{CEM}) becomes site (population) and time dependent. For this reason the system loses
coherence.

It is also worth emphasizing that the nonadiabatic procedure we have described in
the present paper can straightforwardly be realized in a real experiment similar
to those of Refs.~\cite{Orzel_Science291,SF-Diss}.
Therefore, by displacing the contensates with respect to the harmonic trap, as done in the
experiment of Ref.~\cite{Cataliotti_Science293}, and by observing the oscillations
of the center of the atomic density distribution, the effects of the nonstandard
dynamics can be directly observed.


\acknowledgments
I thank the ESF Exchange Grant for support within the activity
``Quantum Degenerate Dilute Systems.''
I also thank G.-L. Oppo and V. Penna for useful discussions.


\hfill



 \end{document}